\documentclass[12pt]{iopart}
\usepackage{graphicx,epsfig,bm,verbatim}

\begin{document}

\title{Spin Star as a Switch for Quantum Networks}

\date{\today}

\author{Man-Hong Yung}

\eads{\mailto{mhyung@chemistry.harvard.edu}}

\address{Department of Chemistry and Chemical Biology, Harvard University, Cambridge MA, 02138, USA}
\address{Department of Physics, University of Illinois at Urbana-Champaign, Urbana IL 61801-3080, USA}

\pacs{03.67.HK, 03.67.Lx, 03.67.-a}

\begin{abstract}
Quantum state transfer is an important task in quantum information processing. It is known that one can engineer the couplings of a one-dimensional spin chain to achieve the goal of perfect state transfer. To leverage the value of these spin chains, a spin star is potentially useful for connecting different parts of a quantum network. In this work, we extend the spin-chain engineering problem to the problems with a topology of a star network.  We show that a permanently coupled spin star can function as a network switch for transferring quantum states selectively from one node to another by varying the local potentials only. Together with one-dimensional chains, this result allows applications of quantum state transfer be applied to more general quantum networks. 

\end{abstract}

\maketitle


\section{Introduction}
Quantum information processing (QIP) with solid state devices is currently a highly challenging, though not impossible, task in quantum computation and quantum simulation \cite{Nielsen2000, Buluta2009, Kassal2010, Yung2010a, Yung2010b}. Due to decoherence \cite{Giulini1996,shor1995}, quantum information stored in physical systems is much more fragile than classical information in terms of fidelity preservation; it is expected that quantum error correction (QEC) \cite{Steane1998} will play an important role to maintain the fidelity of the phase sensitive quantum states. In the standard paradigm of QEC, it is crucial that spatially correlated errors (at least on the same logical qubit) have to be avoided. On the other hand, to minimize thermal noises in solid state systems, it is often preferable to lower the operating temperatures. This, however, would typically increase the ``quantumness" of the noises. For example, indirect interaction between qubits would be induced through the interaction with the environment (for a related recent review, see e.g. \cite{Privman07}). To avoid the ``attacks" of correlated errors, it is conceivably advantageous to encode logical qubits over spatially separated physical qubits. 

To allow the physically separated qubits to communicate with each other, it is necessary that some sort of robust quantum channels are available. This problem is often called the quantum state transfer (QST) problem. Lots of different state transfer schemes \cite{Wojcik05,Greentree06,Feder06,Huo06,Zhou06,Romero06,Lyakhov06,Bayat07,Jahne07,Gong07, YungQIC04} and further development \cite{Perales05,Osborne06,Chen06a,Fitzsimons06,Yung06a,Kay07,Cubitt07,Amico07,Hartmann07, Plastina07, Alvarez2010} have been proposed in recent years (earlier references can be traced from here \cite{Burgarth07}). However, most of them focus only on one-dimensional and single qubit transfer. For this reason, this work is motivated to fill up the gap, or stimulate the development in this direction.

We are interested in the problem of connecting different parts of a quantum network. In classical computing networks, this function is carried out by network hubs and/or switches \cite{hub}. Likewise, quantum switches are conceivably necessary for building up quantum networks. In this work, we show that a group of permanently coupled spins in the star topology \cite{Breuer04,Hutton04a,Hutton04b} can function as a quantum switch by varying the local potentials only. Any quantum state initialized at one spin at the edge (called node) will be transfered to any other node through the natural Hamiltonian evolution. Note that this is not the same as the method of quantum cloning \cite{Chiara05,Chen06b}, as the final state is required to be perfectly transferred to a specific node only.

\begin{figure}[t]
\begin{center}
\scalebox{0.6}{\includegraphics{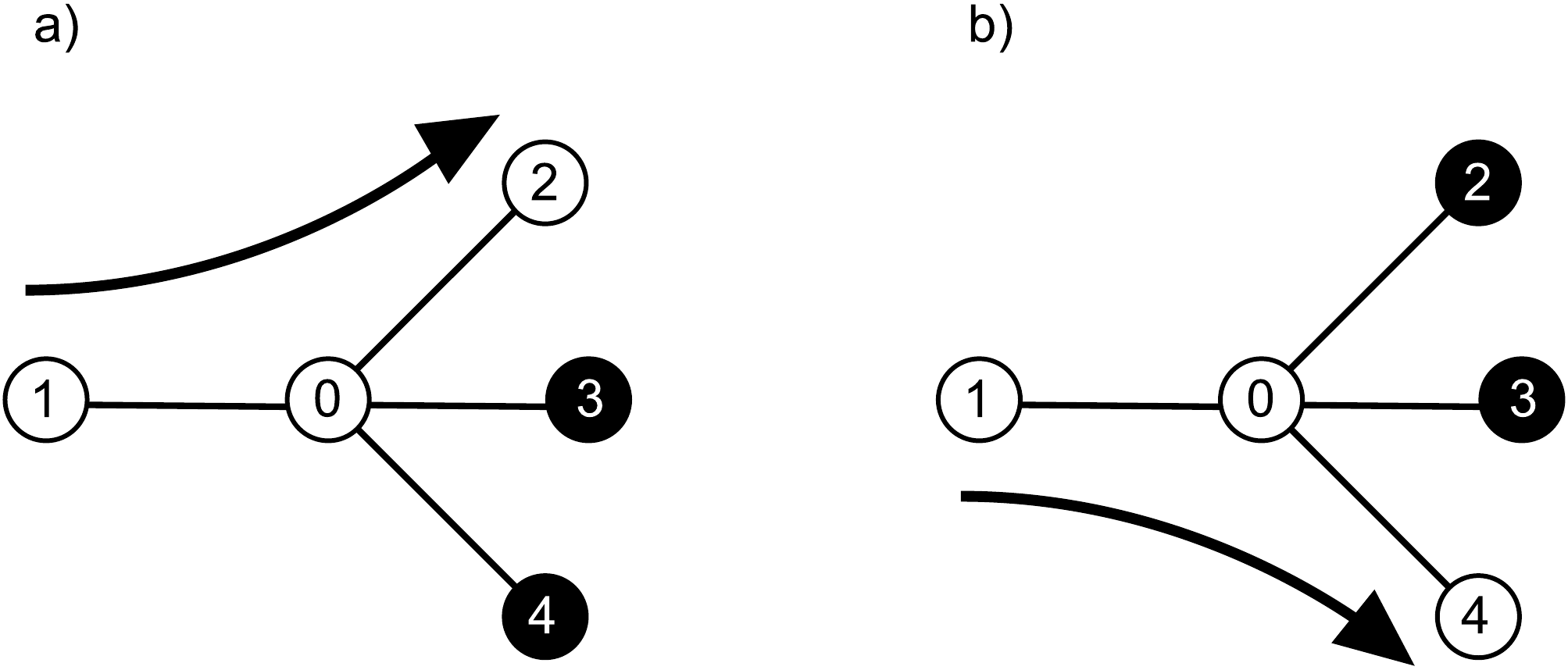}}
\caption{\label{figure} Quantum state transfer with a permanently coupled spin star. Each node may be a port connecting another part of the network. a) By engineering the right set of local potentials, a quantum state can be transfered from node 1 to node 2 under the free Hamiltonian evolution. b) The state can also be selectively transferred to another node by adjusting the local potentials.}
\end{center}
\end{figure}

We note that the solution to this problem is not unique; there are at least two alternatives ways of achieving the same goal. First, if we have perfect controls over the interaction between the spins, then one can simply perform a swap operation from node A to the central node, and then another swap to node B. Second, if the spins are permanently coupled, but strong local pulses are available to effectively control pairwise interactions, as in the case of NMR quantum computing \cite{Jones1998}, then quantum states can be swapped through as well. These two conditions, namely full control and strong local pulses, are not necessary available in many proposals of solid state implementations. Here we assume that we do not have these conditions i.e., spins are permanently coupled and local potential can be varied only mildly (it turns out to scale as $\sqrt{N}$, cf Eq.~(\ref{scaling})). In the following, we shall adopt the approach of spin-chain engineering problem  \cite{Christandl04,Yung04,Yung06,Kost07} which aims at minimize dynamical controls by achieving quantum state transfer with the free Hamiltonian evolution (see Fig. \ref{figure}). It has been \cite{Stockburger07} pointed out that such approach is less affected by the gate control noise. Our strategy is to change the local potentials acting on different sites; similar, but not identical, idea has been explored in a one-dimensional ferromagnetic chain \cite{Plastina07}. 

\section{Definition of the QST problem}
To solve this problem, we need to ``engineer", or design, a suitable Hamiltonian for a system with $N{+}1$ spins ($N$ edge nodes and $1$ central node). Inspired by the spin-chain engineering problem, we consider a Hamiltonian of the following form:
\begin{equation}
H = \frac{{\omega _0 }}{2}\sum\limits_{j = 1}^N {\left( {\sigma _0^x \sigma _j^x  + \sigma _0^y \sigma _j^y } \right)}  + \sum\limits_{j = 0}^N {\frac{{\lambda _j }}{2}\left( {\sigma _j^z  + 1} \right)} \quad .
\end{equation}
Here we assume that the couplings $\omega_0$ between the edge spins and the central spin are the same, but the local potentials $\lambda_j$  could be varied. The $\sigma_j$'s are the standard Pauli matrices acting on the node $j$. With the Hamiltonian $H$, we define 
\begin{equation}
U\left( t  \right) \equiv e^{ - iH t } 
\end{equation}
to be the time evolution operator for a period of time $t$. Note that $\left[ {H,\sigma _z } \right] = 0$, the Hamiltonian $H$ can be block-diagonalized by the basis consisting of fixed number of spin-up's and down's. Therefore, $H$ is spin-conserving. This implies that 
\begin{equation}
U\left| {0_0 0_1 0_2 ...0_N } \right\rangle  = \left| {0_0 0_1 0_2 ...0_N } \right\rangle
\end{equation}
for all times. Suppose at $t=0$, nodes 1 is encoded with an arbitrary qubit state 
\begin{equation}
\alpha \left| 0 \right\rangle  + \beta \left| 1 \right\rangle,
\end{equation}
while the rest are in the definite state $\left| 0 \right\rangle$, the qubit is said to be (perfectly) transfered to node k at $t=\tau$, whenever (up to a phase factor)
\begin{equation}\label{U}
U\left( \tau  \right)\left| {0_0 1_1 0_2 0_3 ...0_k ...0_N } \right\rangle  = \left| {0_0 0_1 0_2 0_3 ...1_k ...0_N } \right\rangle \quad .
\end{equation}

\subsection{Transformation to a four-spin problem}
The question is what values of $\omega_0, \lambda_j$ and $\tau$ would make perfect QST (Eq.(\ref{U})) possible? To answer this question, we note that we can confine ourselves to the single-spin subspace defined by
\begin{equation}
\{ \left| \bm{0}\right\rangle  \equiv \left| {1_0 0_1 0_2 ...0_N } \right\rangle,
\left| \bm{1} \right\rangle  \equiv \left| {0_0 1_1 0_2 ...0_N }\right\rangle, ... \}. 
\end{equation}
In this subsapce, the Hamiltonian $H_S$ is of the arrow-head form:
\begin{equation}
H_S = \left( {\begin{array}{*{20}c}
   {\lambda _0 } & {\omega _0 } & {\omega _0 } &  \cdots  & {\omega _0 }  \\
   {\omega _0 } & {\lambda _1 } & 0 &  \cdots  & 0  \\
   {\omega _0 } & 0 & {\lambda _2 } &  \cdots  & 0  \\
    \vdots  &  \vdots  &  \vdots  &  \ddots  &  \vdots   \\
   {\omega _0 } & 0 & 0 &  \cdots  & {\lambda _N }  \\
\end{array}} \right) \quad .
\end{equation}

Without loss of generality, we consider state transfer from node 1 to node 2. The transition amplitude can be expanded by the following:
\begin{equation}\label{Trans.Amp}
\left\langle \bm{2} \right|U ( \tau ) \left| \textbf{1} \right\rangle  = \sum\limits_{k = 0}^N {\langle \bm{2} | \bm {e}_k \rangle } \left\langle {\bm{e}_k } | \bm {1} \right\rangle e^{ - iE_k \tau } \quad ,
\end{equation}
where $\left| {\bm{e}_k } \right\rangle $ is an eigenvector of $H_S$, and $E_k$ the corresponding eigenvalue. 

For the purpose of performing perfect state transfer, we require
\begin{equation}
\left| {\left\langle \bm 2 \right|U\left| \bm 1 \right\rangle } \right| = 1 \quad.
\end{equation}
Generally, it is in principle possible to choose any complex phase for Eq. (\ref{Trans.Amp}), i.e. $\left\langle \bm 1 \right|U\left| \bm 2 \right\rangle ^*  = \left\langle \bm 2 \right|U\left| \bm 1 \right\rangle  = e^{i\varphi }$. However, for simplicity, we consider the special case
\begin{equation}\label{specialcase}
\left\langle \bm 2 \right|U\left( \tau  \right)\left| \bm 1 \right\rangle  = \left\langle \bm 1 \right|U\left( \tau  \right)\left| \bm 2 \right\rangle  = 1\quad.
\end{equation}
In this way, any state initialized in the state ${\alpha \left| 000...0 \right\rangle  + \beta \left| \bm 1 \right\rangle }$ will be transformed into the state $\alpha \left| 000...0 \right\rangle  + \beta \left| \bm 2 \right\rangle$ under the action of $U(\tau)$. Then, Eq. (\ref{specialcase}) implies that 
\begin{equation}
\left\langle \bm 1 \right|e^{ - i P H_S P\tau } \left| \bm 2 \right\rangle  = \left\langle \bm 1 \right|e^{ - i H_S \tau } \left| \bm 2 \right\rangle \quad,
\end{equation}
where 
\begin{equation}
P \equiv \left| \bm 2 \right\rangle \left\langle \bm 1 \right| + \left| \bm 1 \right\rangle \left\langle \bm 2 \right| + \sum\limits_{k = 3}^N {\left| \bm k \right\rangle \left\langle \bm k \right|}
\end{equation}
is the permutation operator for node 1 and 2. This suggests that $H_S$ is invariant about the exchange of the two spin positions:
\begin{equation}
P H_S P = H_S , 
\end{equation}

In other words, $H_S$ must be symmetrical with respect to node 1 and 2. Therefore, we need to set 
\begin{equation}
\lambda _1  = \lambda _2 \quad .
\end{equation}
Now, suppose we further impose a symmetry condition for the other spins, and set 
\begin{equation}
\lambda _3  = \lambda _4  = ... = \lambda _N \quad,
\end{equation}
then within the subspace $\left\{ {\left| \bm 0 \right\rangle ,\left| \bm 3 \right\rangle_{new} ,\left| \bm 1 \right\rangle ,\left| \bm 2 \right\rangle } \right\}$, where 
\begin{equation}\label{3_to_N}
\left| \bm 3 \right\rangle _{new}  \equiv \left( {\left| \bm 3 \right\rangle _{old}  + \left| \bm 4 \right\rangle  + ... + \left| \bm N \right\rangle } \right)/\sqrt M \quad,
\end{equation}
The Hamiltonian is reduced into a $4\times 4$ Hamiltonian $H_4$ in the following form:
\begin{equation}\label{H_4_define}
H_4  = \left( {\begin{array}{*{20}c}
   a & b & c & c  \\
   b & d & 0 & 0  \\
   c & 0 & e & 0  \\
   c & 0 & 0 & e  \\
\end{array}} \right) \quad,
\end{equation}
where 
\begin{eqnarray}
a \equiv \lambda_0 \quad, \\
b \equiv \sqrt {M} \omega _0 \quad, \nonumber \\
c \equiv \omega _0 \quad, \nonumber \\
d \equiv \lambda _3  = \lambda _4  = ... = \lambda _N \quad , \nonumber \\
e \equiv \lambda _1  = \lambda _2 \quad. \nonumber
\end{eqnarray}
This fact can be readily checked by direct multiplication. Explicitly,
\begin{eqnarray}
H\left| \bm 0 \right\rangle  = \lambda _0 \left| \bm 0 \right\rangle  + \sum\limits_{k = 3}^N {\omega _0 } \left| \bm k \right\rangle  + \omega _0 \left| \bm 1 \right\rangle  + \omega _0 \left| \bm 2 \right\rangle  = a\left| \bm 0 \right\rangle  + b\left| \bm 3 \right\rangle  + c\left| \bm 1 \right\rangle  + c\left| \bm 2 \right\rangle \, , \nonumber  \\
H\left| \bm 3 \right\rangle  = \sqrt M \omega _0 \left| \bm 0 \right\rangle  + \lambda_3 \left| \bm 3 \right\rangle  = b\left| \bm 0 \right\rangle  + d\left| \bm 3 \right\rangle \, , \nonumber \\
H\left| \bm 1 \right\rangle  = \omega _0 \left| \bm 0 \right\rangle  + \lambda _1 \left| \bm 1 \right\rangle  = c\left| \bm 0 \right\rangle  + e\left| \bm 1 \right\rangle \, , \nonumber \\
H\left| \bm 2 \right\rangle  = \omega _0 \left| \bm 0 \right\rangle  + \lambda _2 \left| \bm 2 \right\rangle  = c\left| 0 \right\rangle  + e\left| \bm 2 \right\rangle \, .
\end{eqnarray}

Therefore, the $M \equiv N-2$ edge spins act effectively as a single ``renormalized" spin, with 
\begin{equation}
\omega _0  \quad \to \quad \omega _0 \sqrt {M} 
\end{equation}
but local potentials $\lambda_k$ ($k\ge
 3$) unchanged. The many-spin problem is then reduced to a four-spin problem; a very similar procedure was  described in the work of Burgarth and Bose \cite{Burgarth2006}.

The transition amplitude $\left\langle \bm{2} \right|U ( \tau ) \left| \textbf{1} \right\rangle$ is still given by the form of Eq. (\ref{Trans.Amp}) provided that we now consider $\left| \bm{e}_k \right\rangle $ and $E_k$ as the eigenvectors and eigenvalues of $H_4$ and sum up to $N \to 3$, and adopt the new basis  $\left\{ {\left| \bm 0 \right\rangle ,\left| \bm 3 \right\rangle_{new} ,\left| \bm 1 \right\rangle ,\left| \bm 2 \right\rangle } \right\}$.

\subsection{Exchange symmetry}
Here we shall further exploit the advantages gained from the exchange symmetry. Since by definition 
\begin{equation}
P^2  = P \quad,
\end{equation}
the requirement 
\begin{equation}
P H_4 P = H_4
\end{equation}
or equivalently $\left[ {P,H_4 } \right] = 0$ implies that the eigenvectors of $H_4$ are also eigenvectors of $P$, with eigenvalues $\pm 1$. Consider explicitly the eigenvalue equation 
\begin{equation}
H_4 \left| {\bm {e}_k } \right\rangle  = E_k \left| {\bm{e}_k } \right\rangle \quad.
\end{equation}
By direct substitution, the vector $\left( {0,0,1, - 1} \right)^T$ is an eigenvector of $H_4$ with eigenvalue $e$. By inspection, it is also an eigenvector of $P$ with eigenvalue $-1$.

 For $E_k \ne e$, we shall assume that the exchange of 1 and 2 gives no phase change. In other words,  
\begin{eqnarray}\label{P}
 P\left| {\bm {e}_k } \right\rangle  &=&  - 1\left| {\bm {e}_k } \right\rangle \quad {\rm{for}}\quad E_k  = e \quad , \\ 
 P\left| {\bm {e}_k } \right\rangle  &=&  + 1\left| {\bm {e}_k } \right\rangle \quad {\rm {for}} \quad E_k  \ne e  \nonumber \quad .
 \end{eqnarray}
Consequently,  Eq. (\ref{Trans.Amp}) becomes
\begin{equation}\label{plus_minus}
\left\langle \bm 2 \right|U\left( \tau  \right)\left| \bm 1 \right\rangle  = \sum\limits_{k = 0}^N {\left\langle \bm 1 \right|\left( {P\left| {\bm e_k } \right\rangle } \right)} \left\langle {\bm e_k } \right.\left| \bm 1 \right\rangle e^{ - iE_k \tau }  \quad.
\end{equation}
To satisfy the perfect state transfer condition, or $\left\langle \bm 2 \right|U\left( \tau  \right)\left| \bm 1 \right\rangle  = 1$. It is necessary that the phase factors $e^{ - iE_k \tau } $ can cancel the ${\pm 1}$ phase according to Eq. (\ref{P}).

It is remarkable that this property is highly analogous to that \cite{Yung04} in the 1-D spin-chain engineering problem. In fact, with the knowledge of the eigenvalues and the exchange properties of the eigenvectors, one can determine the matrix elements of the Hamiltonian $H_4$. This type of problem is called \textit{inverse eigenvalue problem} \cite{Gladwell1986}. 

\section{Mapping to an inverse eigenvalue problem}
Inverse eigenvalues problems are often more challenging than the ordinary eigenvalue problems. In particular, not many analytic solutions have been obtained in the spin-chain engineering problem. Fortunately, because our $(N+1)$-spin problem is now effectively a four-spin problem, the inverse eigenvalue problem is exactly solvable.

With Eq. (\ref{P}), to ensure we have $\left\langle \bm 2 \right|U\left( \tau  \right)\left| \bm 1 \right\rangle = 1$ in Eq. (\ref{Trans.Amp}), consider Eq. (\ref{plus_minus}). The goal is to choose the phase factors $e^{ - iE_k \tau }$ to cancel the phase factors $\pm 1$ due to the exchange operator $P$. We set for $E_k = e$, 
\begin{equation}
e\tau  = \pi \quad,
\end{equation}
and for $E_k  \ne e$, we set $E_k \tau  =0, \pm 2 n \pi$, where $n$ is a positive integer. This can be achieved by choosing the following eigenvalue spectrum 
\begin{equation}
\left\{ {0,e, \pm \eta e} \right\} \quad.
\end{equation}
Given any 
\begin{equation}
\eta = 2 n \quad,
\end{equation}
if we know a solution of $e$ satisfying Eq. (\ref{equations}) for the Hamiltonian $H_4$, then at 
\begin{equation}
t=\tau=\pi / e \quad,
\end{equation}
combining Eq. (\ref{P}), Eq. (\ref{Trans.Amp}) and the conditions above, we have $\left\langle \bm 2 \right|U\left( {\pi /e} \right)\left| \bm 1 \right\rangle = 1$ and the state transfer problem can be solved. 

\section{Solution to the inverse eigenvalue problem}
We shall briefly describe how the solutions (i.e., to determine a set of $\{a,b,c,d,e\}$) to the inverse eigenvalue problem can be found and skip technical details. 
To get started, consider the characteristic polynomial of $H_4$:
\begin{equation}
p\left( \lambda  \right) \equiv \det \left| {H_4  - \lambda } \right|/\left( {e - \lambda } \right) \quad,
\end{equation}
where $\det \left| {H_4 - \lambda I} \right| = \left( {e - \lambda } \right)\left[ {\left( {a - \lambda } \right)\left( {d - \lambda } \right)\left( {e - \lambda } \right) - b^2 \left( {e - \lambda } \right) - 2c^2 \left( {d - \lambda } \right)} \right]$. Here $\lambda$ is the variable of the function $p(\lambda)$. To simplify the notation, we write 
\begin{equation}\label{p_equations}
p\left( \lambda  \right) = \Lambda _0 e^3  + \Lambda _1 e^2 \lambda  + \Lambda _2 e\lambda^2  - \lambda ^3,
\end{equation}
where 
\begin{eqnarray}\label{equations}
 e^3 \Lambda _0   &=& ade - b^2 e - 2c^2 d \quad ,\\ 
 e^2 \Lambda _1   &=& b^2  + 2c^2  - ad - \left( {a + d} \right)e \nonumber \quad ,\\ 
 e \, \Lambda _2  &=& a + d + e \quad .\nonumber
\end{eqnarray}

With the eigenvalue spectrum chosen to be $\left\{ {0,e, \pm \eta e} \right\}$, where $\eta$ is a tuning parameter. We also have $\det \left| {H_4  - \lambda } \right| = \lambda \left( {e - \lambda } \right)\left( {\eta ^2 e^2  - \lambda ^2 } \right) = \left( {e - \lambda } \right)\left( {\eta ^2 e^2 \lambda  - \lambda ^3 } \right)$. Comparing it with Eq. (\ref{p_equations}), we have
\begin{equation}
\Lambda _0  = \Lambda _2  = 0
\end{equation}
and 
\begin{equation}
\Lambda _1  = \eta ^2 \quad.
\end{equation}
For simplicity, we measure all quantities in unit of $c$, or equivalently set $c=1$ and hence 
\begin{equation}
b^2  = M \quad, 
\end{equation}
which means that $b$ and $c$ is nowdetermined. 

Eliminating $a$ and $d$ in Eq. (\ref{equations}), we have 
\begin{equation}
g_M \left( {e;\eta } \right) \equiv x_0  + x_2 e^2  + x_4 e^4  + x_6 e^6  = 0 \quad ,
\end{equation}
where 
\begin{eqnarray}
x_0  &=& M + 2  \quad,\\
x_2  &=& 3 - \eta ^2 \quad , \nonumber \\
x_4  &=& \left( {3/2} \right)\left( {1 - \eta ^2 } \right) \quad , \nonumber \\
x_6  &=& \left( {1/4} \right)\left( {1 - \eta ^2 } \right)^2 \nonumber \quad .
\end{eqnarray}

Given $\eta$ and $M$, as we have fixed $b$ and $c$, our goal is to find $e$ (and hence $a$ and $d$) which solves Eq. (\ref{equations}). This can be done efficiently using any numerical method. The key point here is that the complexity for solving the problem does not scale with the number of spins $N$ involved. 

\subsection{Examples}
\begin{figure}[htb]
\begin{center}
\scalebox{0.6}{\includegraphics{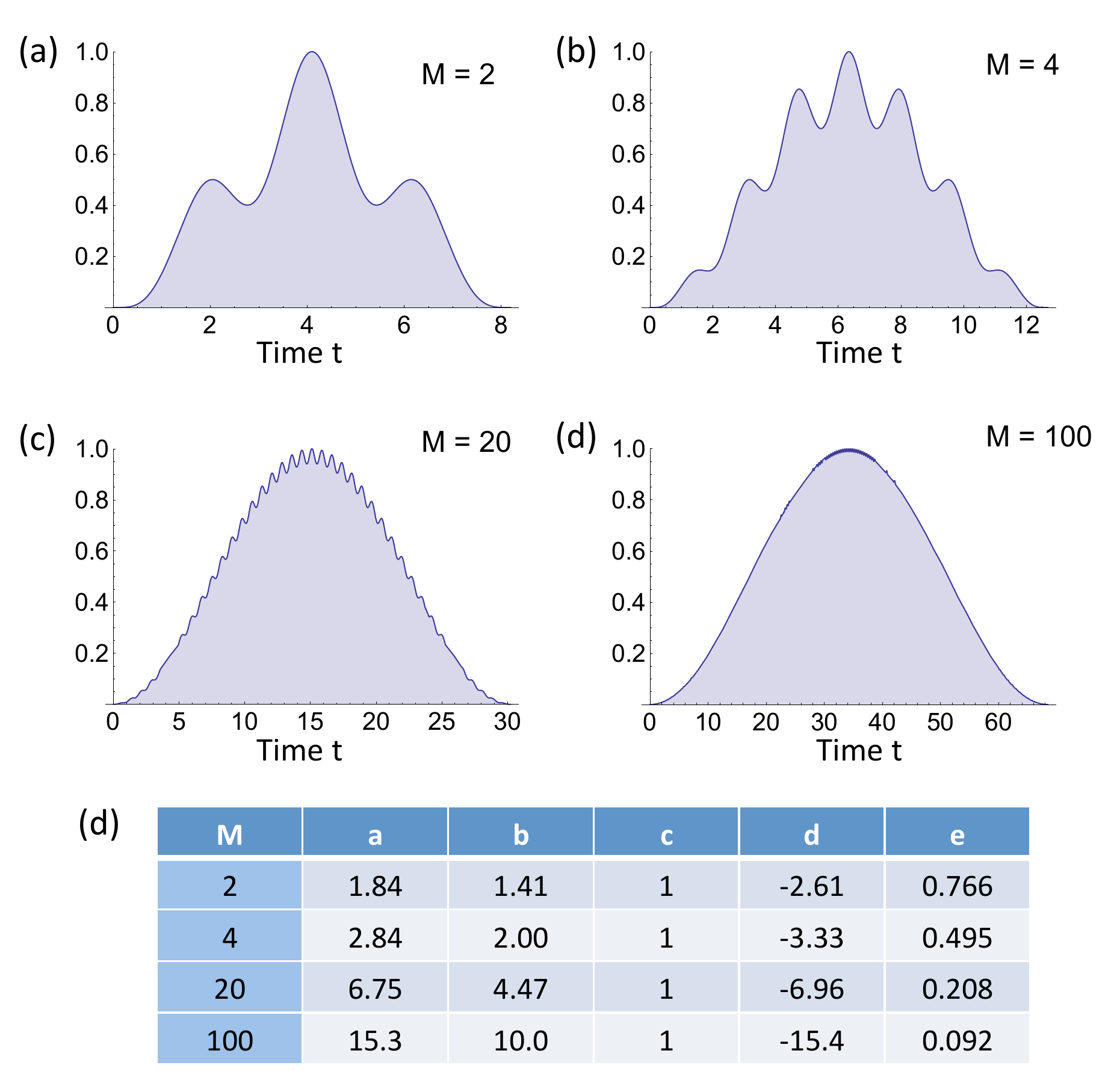}}
\caption{\label{figure2} Quantum state transfer with a permanently coupled spin star of $M+2+1$ spins, where $M$ is the number of spins get normalized (see Eq. (\ref{3_to_N})). The vertical axis is $F(t) \equiv \left| {\left\langle {\bm 2} \right|U\left( t \right)\left| {\bm 1} \right\rangle } \right|^2$, the transfer fidelity; it reaches the maximum ($F=1$) when $t=\pi/e$, which corresponds to a perfect state transfer from state ${\left| {\bm 1} \right\rangle }$ to ${\left| {\bm 2} \right\rangle }$. Figure (a)-(d) show the cases of different values of $M$. Figure (d) is a table listing the corresponding matrix elements of the Hamiltonian defined in Eq. (\ref{H_4_define}).  }
\end{center}
\end{figure}

As an example, for $M=2$ (i.e., 4+1 spins), we can choose $\eta = 4 $ (i.e., with eigenvalue spectrum $\left\{ {0,e, \pm 4e} \right\}$), one of the solutions is $e=-d=0.516$, $a=0$ and of course $c=1$ and $b=\sqrt 2 $. Note that the local potentials $e$ and $d$ are roughly the same order as the couplings $c$. Consequently, a qubit of information encoded at node 1 at time $t=0$ will evolve to node 2 at $t=\pi/e$ under the free evolution of this Hamiltonian, and the problem is solved. The fidelity as a function of time is shown in Figure \ref{figure2}.

Suppose we now want to transfer the state from node 1 to node 3 instead of 2, as the couplings $\omega_0$ are all the same, we just need to exchange the local potential between 2 and 3, $\lambda _2  \leftrightarrow \lambda _3 $. Practically, we may apply a global offset field, so that only the ``active" nodes are subject to some non-zero potentials, and all other $\lambda _k  = 0$ (or some constant value).

\subsection{Scaling Analysis}
Lastly, to complete our discussion, we should also consider the constraints of $\eta$ which at this point seems to be arbitrary. In fact, to ensure (a pair of) real roots of $g_M\left( {e;\eta } \right)$ exist, we require the global minimum of $g_M\left( {e_*;\eta } \right)$ (denote the corresponding value of $e$ as $e_*$), as a function of $e$, to be less than zero. We found 
\begin{equation}\label{e_star}
e_*^2  = \left( {6 + 2\sqrt 3 \eta } \right)/3\left( {\eta ^2  - 1} \right)
\end{equation}
which suggests that $\eta  > 1$. The condition $g_M \left( {e_* ;\eta } \right) < 0$ implies that 
\begin{equation}
\frac{{4\sqrt 3 }}{9}\frac{{\eta ^3 }}{{\eta ^2  - 1}} > N \quad.
\end{equation}
For large $N$, $\eta$ needs to be at least of order $O(N)$. Therefore, from Eq. (\ref{e_star}), $e$ is of order $O(1/\sqrt{M})$. In this limit, from Eq. (\ref{equations}), we have the local fields 
\begin{equation}\label{scaling}
a \approx -d \approx O(\sqrt{M}).
\end{equation}
All of these are in agreement with the numerical results in Fig. \ref{figure2}.

In conclusion, we have demonstrated how the spin-chain engineering problem can be generalized to the topology of spin-star, which can be potentially useful to function as a switching device for building up quantum networks.  Our main idea is the reduction of the many-spin problem to a four-spin problem which is exactly solvable. Together with one-dimensional chains, this result allows the quantum state transfer to more general quantum networks, e.g. consisting of superconducting devices \cite{Hanneke2010}.



\ack
M.H.Y. are grateful to the funding sources: NSF grant EIA-01-21568 and Croucher Foundation.




\begin{thebibliography}{99}
\bibitem{hub}
In classical computing networks, hubs transmit packets to all of the ports, while switches forward packets to a specific port only. Since quantum information cannot be (perfectly) cloned, it is more appropriate to refer the spin-star as a quantum switch.
\end{thebibliography}
\end{document}